\documentclass[twocolumn,showpacs,preprintnumbers,amsmath,amssymb]{revtex4}
\usepackage{graphicx}% Include figure files
\usepackage{dcolumn}% Align table columns on decimal point
\usepackage{bm}% bold math

\begin{document}
\preprint{}
\title{The Diagnostics of the Chemical Reaction Zone \\ at the Detonation of Condensed Explosives}
\author{Nataliya P. Satonkina}
\altaffiliation{Lavrentyev Institute of Hydrodynamics, Novosibirsk, 630090 Russia\\
Novosibirsk State University, Novosibirsk, 630090 Russia}%
%\altaffiliation[Also at]{{Lavrentyev Institute of Hydrodynamics, Novosibirsk, 630090 Russia\\Novosibirsk State University, Novosibirsk, 630090 Russia}%
 \email{snp@hydro.nsc.ru}
\date{\today}
\newcommand\revtex{REV\TeX}

\begin{abstract}
The highly-sensitive method is proposed for the real-time diagnostics of the chemical peak (von
Neumann peak) at detonation of brisant high explosives. The absence of the direct link between the
pressure and the course of chemical reactions was shown. For TNT, the influence of the structure of
charge on the kinetics of chemical peak was demonstrated.
\end{abstract}

\pacs{82.33.Vx, 82.40.Fp, 82.60.Qr}

\maketitle

%\noindent

\section{  Introduction}

According to the Zeldovich -- von Neumann -- D\"{o}ring (ZND) theory, detonation wave consists of
the thin (several intermolecular distances) shock wave region, the chemical reaction zone, the
Chapman -- Jouguet plane (CJ point in one-dimensional case), and the Taylor rarefaction wave. It is
commonly accepted that the CJ point which divided the regions of subsonic and supersonic flows
corresponds to the ending of the chemical reaction zone. However, a strict validation of the link
between the ending of the chemical peak and the CJ point is absent. Since the chemical peak in the
ZND model is defined as a region of high pressure, it is investigated solely using the time
dependencies of the pressure obtained directly or indirectly by the measurement of mass velocity
and temperature. Many authors note the difficulty to set the CJ point which is explained, for
example, by an incomplete chemical reaction.

Data on the duration of the chemical peak show large scatter (for a review see \cite{loboiko}) as
well as the different dependence on the charge configuration, density, length and diameter. This
could be due to the absence of the universal way to find the singularity (the kink point) at the
pressure profile, and also it indicates the methodical difficulties.

Neither density nor pressure give the direct information on the chemical composition of the matter.
Zeldovich and Kompaneets noted in their book on the detonation theory \cite{zeldkomp} that the
detonation velocity, the mass velocity of the explosion products, and the pressure do not depend on
the speed of the chemical reaction. This means the absence of the relation between the duration of
chemical reaction and mechanic parameters.
% if the reaction mechanism ensures the tangent condition of the Michelson line and the Hugoniot adiabata.

As shown in the works \cite{fedorov99,dremin,titov,pccp15,satonkinajap,satonkinafgv16}, high
electric conductivity at the detonation of condensed HEs with negative oxygen balance is related to
the presence of carbon. In the works \cite{satonkinajap,satonkinafgv16}, the link between the value
of conductivity and the mass fraction of carbon is shown for the whole detonation wave. Based on
these results, we propose in this paper a new method for the diagnostics of the state of the matter
in the von Neumann peak based on the electric conductiovity profile.

{\bf Results of the investigation of the duration of the pressure peak}

The duration of the chemical peak was investigated earlier in the works
\cite{loboiko,fedorov99,dremin,titov}.

The increase of the chemical peak duration $\Delta t$ with an increase of the TATB charge diameter
was obtained in the work of Loboiko \cite{loboiko} using the registration of the light emission
from the shock wave in chloroform. There was also an influence of the charge length, for RDX,
$\Delta t$ increased with the increase of the length at constant diameter. Authors note that there
is no clearly defined boundary between the reaction zone and the Taylor wave. They interpret this
as an indirect evidence of the incompleteness of chemical reactions. The duration of the chemical
peak was determined using the averaging of pressure profiles from several experiments followed by
the smoothing and takin the logarithm. This procedure gave two staright lines in semilogarithmic
cordinates. The intersection of these lines was taken as the CJ point.

In the work of Fedorov \cite{fedorov99} , the detonation parameters and the density profiles for
individual and mixed HEs were obtained by the Fabry -- P\'{e}rot method with the nanosecond
resolution. From thiese profiles, the pressure in the von Neumann peak, the CJ point and the
duration of the chemical peak were determined. Besides the common pressure profiles with a kink
corresponding to the CJ point, the smooth profiles were obtained as well as the nonmonotonic
profiles with two kinks. For HEs of powder density, profiles with large oscillations caused by the
hihg degree of nonuniformity were also observed. The duration of the chemical peak increased with
the increase of the charge size.

In the work of Dremin\cite{dremin}, the increase of $\Delta t$ with the decrease of density was
obtained for TNT and RDX by the magnetoelectric method of the mass velocity registration. For
RDX\cite{dremin}, the duration increased from 100 ns for the density of $\rho=1.72$ g/cm$^3$ to 400
ns ($\rho=0.95$ g/cm$^3$). Less pronounced increase of the duration by 20\% with the increase of
the charge diameter from 18 mm to 40 mm with the fixed density of $\rho=1.00$ g/cm$^3$ was also
observed.

Results of recent works listed are summarized in the Table 1. Maximum and minimum values for PETN
differ by the factor 16. The paper \cite{doronin} contain the data for several HEs. The values
obtained for the same density by different methods differ by several times.

In the work of Titov\cite{titov}, the size of nanodiamonds obtained in the explosion of cylindrical
and spherical charges of the same composition with different initiation methods was investigated.
No dependence was found, which leaded to the conclusion that nanodiamods form in the chemical peak
with parameters independent on the geometry.

\begin{center}
\begin{table}
\caption{Data on the duration of the chemical peak} \label{temptab}
  \begin{tabular}{c c c c }\hline
   HE       & $\rho$, g/cm$^3$  & $\Delta t$, ns      & ref \\  \hline
TNT press.1 & 1.53      & 190         & \cite{loboiko}   \\
TNT press.2 & 1.62      & 330         & \cite{loboiko}   \\
 TNT cast   & 1.56      & 290         & \cite{loboiko}   \\
 TNT        & 1.62      & 305             &  \cite{fedorov99}     \\
 TNT        & 1.62      & 260$\div$360    &  \cite{dremin}     \\
 RDX        & 1.44      & 200             & \cite{dremin}   \\
 RDX        & 1.68      & 50$\div$70      & \cite{loboiko}           \\
 RDX        & 1.72      & $\leq$100       &  \cite{dremin}      \\
 PETN       & 1.73      & 80          & \cite{loboiko}   \\
 PETN       & 1.74      & <5          & \cite{loboiko}   \\
 PETN       &           & 25          & \cite{pccp15}    \\
 HMX        & 1.86      & 60          & \cite{loboiko}   \\
 HMX        & 1.90      & 40          & \cite{loboiko}   \\
 HMX        &           & 30          & \cite{pccp15}    \\

\hline
\end{tabular}
\end{table}
\end{center}

Thus, there is no common pattern in the influence of the HE density on $\Delta t$ with a visible
link to the charge geometry and configuration,
the presence and the sort on a shell. The large scatter of data in observed related
primarily to the method of determining the CJ point.

  \begin{figure}[h!]
\includegraphics[width=85mm]{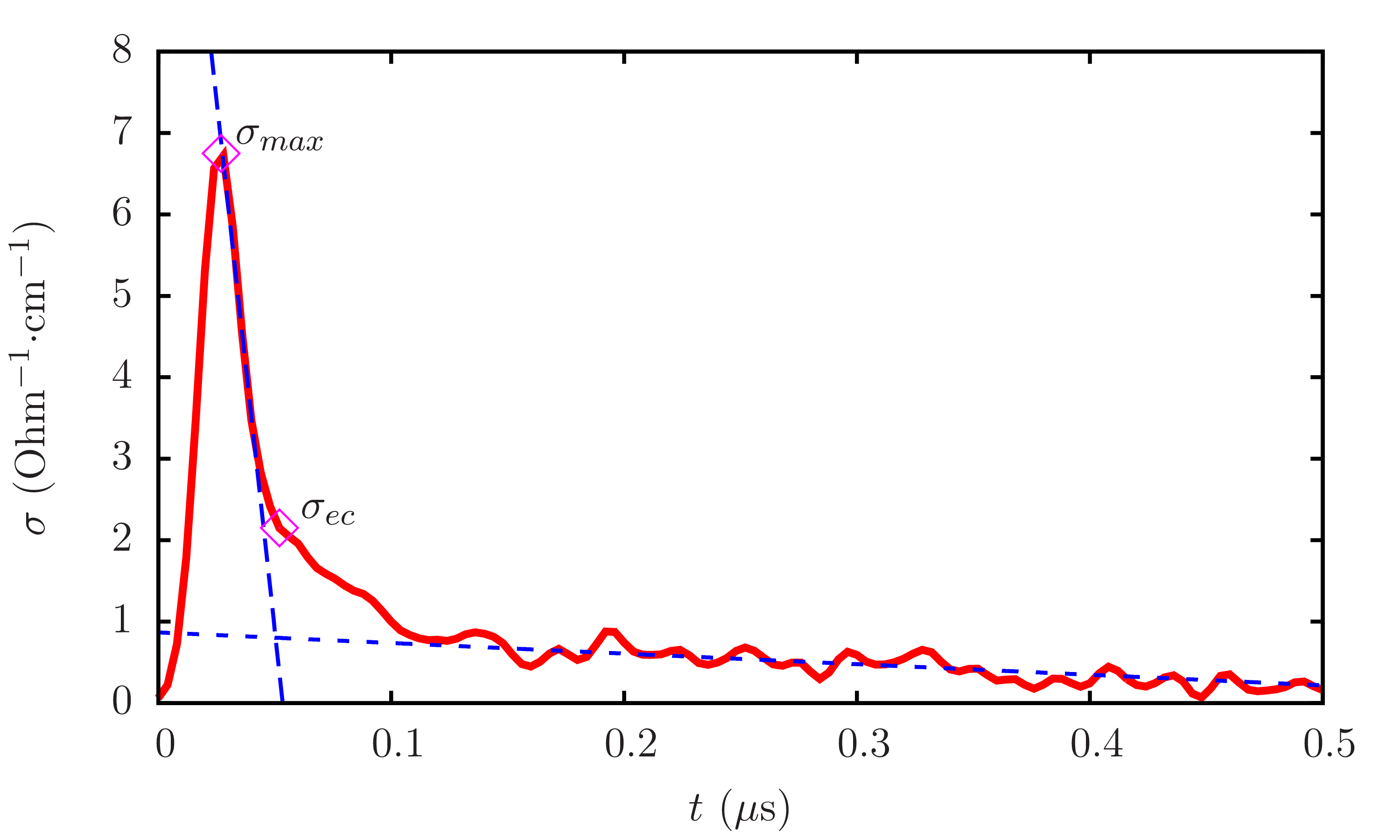} \\
\caption{Conductivity profile at the detonation of condensed HE.}
\end{figure}

{\bf  Methodology}

A typical conductivity profile $\sigma(t)$ at the detonation of condensed HEs with slightly
negative oxygen balance is shown in Fig. 1. The conductivity increases to the maximum value
$\sigma_{max}$ in several nanoseconds, and later decreases until the point labelled $\sigma_{ec}$.
The gradient of this decrease depends on the HE. The method for the determining the duration of the
conductivity peak by an intersection of two approximating straight lines is shown. In the papers
\cite{satonkinajap,satonkinafgv16}, the value of $\sigma_{max}$ is related to the total carbon
content in the HE, and the value of $\sigma_{ec}$ is related to the content of the condensed carbon
which is calculated using the BKW equation (the data of \cite{tanaka}) for the CJ point under the
assumption that the CJ point coincides with the ending of chemical reactions. The given dependence
of the conductivity  $\sigma$ on the carbon content for the mass fraction of carbon from 0 to 0.37
is universal for five HEs.\cite{satonkinajap,satonkinafgv16} This indicates the crucial role of the
carbon for the conductivity process in the chemical peak and in the whole detonation wave.

We assume that the high conductivity at the detonation is provided by the contact mechanism along
the connected highly-conductive carbon nanostructures. The conductivity of carbon varies in a wide
range from 240 to 1250 Ohm$^{-1}$cm$^{-1}$. The condutivity of highly oriented graphite can be as
high as $2\cdot 10^4$ Ohm$^{-1}$cm$^{-1}$ \cite{bib:korobenko99,bib:korobenko03}.

In such HEs as PETN, HMX, RDX, the mass fraction of carbon in the molecule is higher than 0.15
which is sufficient for the formation of elongated structures. In the region where the chemical
reaction ends, the carbon fraction is lower than 0.07 which is not sufficient for contact
conductivity as shown in \cite{symp2014}. Because of such difference, the reaction zone is
distinctly pronounced in the conductivity graph. In carbon-rich HEs such as TNT and TATB, the
amount of carbon is sufficient for the formation of "wires" \ both in the maximum and in the point
of $\sigma_{ec}$. Therefore, the chemical peak is resolved only at the decrease of the density to
powder values \cite{safonov,rubtsov,ershov09}.

The existence of carbon nets is confirmed in following sources. In the work \cite{bib:breusov}, it
was shown that the formation of the carbon structures can not proceed through the intermediate
release of a free carbon. A mechanism of nanodiamod formation was proposed related to the partial
breaking of molecular bonds and the formation and growth of carbon framework. The author of
\cite{bib:anisichkin94,bib:anisichkin07} based on the experiments with isotope label claims that
the oxidation of carbon occurs later that the formation of carbon particles, and confirms fast
clustering of carbon atoms. This agree with the data listed. In the works of Gilev
\cite{gilev02,gilevdis}, it was shown that high maximal conductivity of 250 Ohm$^{-1}$cm$^{-1}$ at
the detonation of TNT can not be explained by the percolation conductivity because even the total
carbon content is insufficient, and the elongated structures with almost metallic conductivity
should be formed. The carbon clustering at heating of molecules of TATB, HMX and RDX was
investigated in \cite{pccp15} using the molecular dynamics method. The carbon nanostructures was
obtained which agrees well with their presence already in the reaction zone. The conclusions were
made that the higher content of carbon decreases the HE sensitivity and increases the stability of
a carbon cluster which hinders the chemical reaction and increases its duration.

The additional indirect confirmation of the presence of elongated "wires" \ are the observation of
the formation of carbon structures under different conditions
\cite{str1,str2,bib:pena,bib:pena1,bib:pinaev} and the carbon filaments found in the detonation
products \cite{vol1,vol2,vol3}.

Thus, the electric conductivity traces the state of carbon. Before the arrival of the detonation
front, HE is dielectric, and all carbon is bound in molecules. The conductivity in the shock front
is insignificant. The peak region (Fig. 1) can be divided into three parts. The conductivity
increase to the maximum corresponds to the decay of initial HE molecules accompanied by the growth
of carbon structures. The oxidation of carbon occurs between the points $\sigma_{max}$ and
$\sigma_{ec}$ leading to the thinning and partial breaking of conductive structures which results
in the decrease of the conductivity. The transition region between 0.05 and 0.1 $\mu s$ could be
caused by the decrease of the intensity of oxidation reactions due to the decrease of
concentrations of reactive components. This transition region is observed on the profiles for all
HEs at high density, and it is weakly pronounced at powder densities. The region of $t>0.1$ $\mu$s
corresponds to the Taylor wave.

We define the ending of the chemical peak as the place within the detonation wave where the decay
of initial HE molecules is finished, and the synthesis of main products which constitute more than
97\% of the detonation products (C, CO, CO$_2$, N$_2$, H$_2$O) already occurs. This point
corresponds to the place of $\sigma_{ec}$ in Fig. 1. The change of the chemical
composition\cite{tanaka} due to the pressure decrease and the shift of the chemical equilibrium
(which is formally also a chemical reaction) corresponds to the Taylor wave.

 {\bf Results of investigation of electric conductivity }

The experimental data used for the analysis were obtained by the high-resolution scheme described
in detail in the work of Ershov\cite{ershov07}. HE was placed inside a thick-walled copper cylinder
with inner diameter of 8 mm. Such shell decreases the critical diameter by 6--8 times
\cite{kobylkin} ensuring the steady detonation. The diameter used was much larger than the critical
one for RDX, PETN and HMX.

The data on the duration of the conductivity peak for RDX, PETN and HMX at different densities are
shown in Fig. 2. The scatter of values is due to the high sensitivity of the method. The maximum
difference of the front arrival related to the constructive features of the scheme used is 0.27 mm
which corresponds to $\sim 50$ ns. Therefore, there is an inherent probability of the widening of
the signal peak. The difference between maximum and minimum values for densities larger than 1.4
g/cm$^3$ corresponds to this estimate, hence, the minimal obtained values should be considered. One
can claim that there is no noticeable density dependence of the conductivity peak duration (the
influence of the density on the chemical peak duration). Since the data shown in Fig. 2 were
obtained at the same geometric charge parameters for different densities, this indicates to the
absence of direct link between the density profile and the chemical reaction zone.

At powder densities ($\rho<$1.4 g/cm$^3$, data of Ershov\cite{ershov10}), a fundamental change of
the conductivity profile occurs at a certain grain size specific for the type HE. The oscillation
arise with a spatial size four times smaller than the grain size. Such correlations were observed
for  PETN (260 and 80 $\mu m$) and for HMX (460 $\mu m$). The profiles for HMX (21 $\mu m$) and RDX
(160 and 11 $\mu m$) are qualitatively the same as for higher densities. The peculiarities for
low-density HEs were observed in the pressure profiles for HMX in the work of
Fedorov\cite{fedorov05} and for PETN in the work Ershov\cite{ershov16}. This was related to the
non-classical detonation regime of the detonation propagation driven by jets. As a result, the
front is highly curved, there is the reacted as well as the initial matter within the same
cross-section. This influence measurements of electric conductivity because electric field force
lines become curved, and the averaging is performed based on the minimization of the electric
resistance rather than along the cross-section. Nevertheless, the influence of the grain size on
the reaction zone was registered by the electric conductivity method. The decrease of the
conductivity peak duration with the decrease of the grain size was observed in \cite{ershov10}
which first of all reflects the influence of the degree of non-uniformity of the initial charge
structure.

\begin{figure}[h!]
\includegraphics[width=85mm]{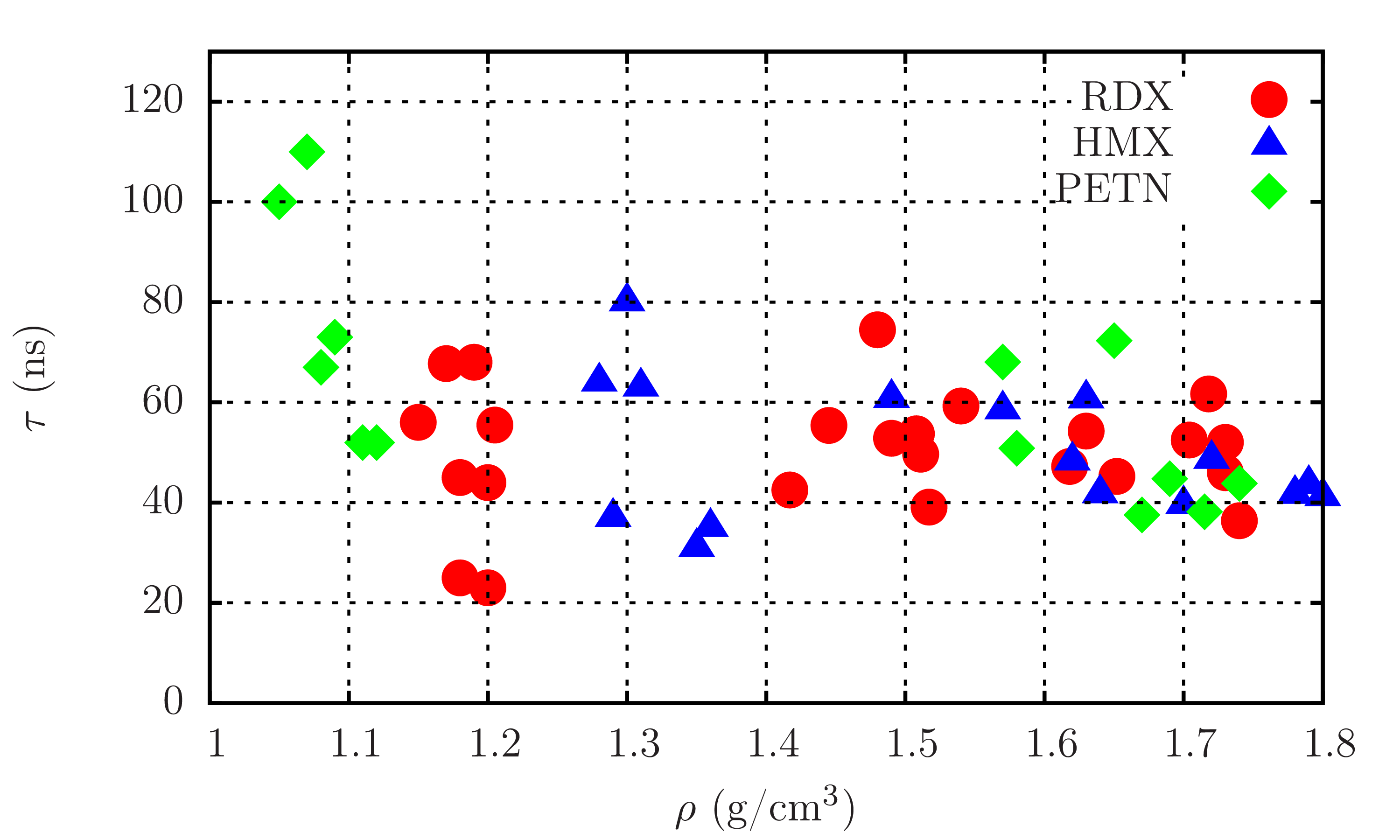} \\
\caption{The duration of the conductivity peak at the detonation of RDX, HMX at PETN at different
density, data \cite{ershov07,ershov10}.}
\end{figure}

%A weak influence of the density on the conductivity peak duration is visible.

The kinetics of chemical reactions should be independent on the charge diameter and length. The
charge geometry however influence the rarefaction. The data on the duration of the chemical peak
obtained from the pressure profile correspond to the decrease of thermodynamic parameters, whereas
the duration of the reaction zone is governed solely by the chemical reaction of the ignition and
burning from initial points related to the hot spot theory. This can explain the independence on
the density.

Considering electric conductivity to be due to carbon nanostructures and taking into account that
the maximum value is connected with the total carbon content, and the value in the CJ point -- to
the condensed carbon content, we get a tool to investigate the chemical reaction zone. The duration
of the conductivity peak defines the duration of the chemical reaction zone. Note that the width of
the conductivity peak increases with the increase of the density and, correspondingly, of the
detonation velocity.

{\bf Diagnostics of the chemical reaction at the detonation of TNT}

Different parameters for cast and pressed TNT are noted in literature. Different duration of the
pressure peak (see Table 1) and the three- to tenfold difference of the critical diameter is
related to different character of nonuniformities dependent of the details of charge preparation.

It is commonly assumed the the hot spots are generated at the nonuniformities, and the chemical
reaction proceeds from these spots. The strong influence of the concentration of hot spots on the
reaction time was shown in the work \cite{fedorov12}. For HMS and PETN, the duration of the
pressure peak in single crystals was several times larger comparing to smaller densities. The role
of hot spots in TNT can be traced using conductivity profiles.

The maximum conductivity of $\sim$25 Ohm$^{-1}$cm$^{-1}$ at the detonation of cast TNT was obtained
in the work \cite{ershov00}. For TNT pressed to the density of 1.57 g/cm$^3$, the conductivity of
100 Ohm$^{-1}$cm$^{-1}$ was obtained in \cite{ershov09}. The values of conductivity for pressed TNT
is three times higher whereas the values of detonation velocity, pressure and density differ only
slightly. The character of conductivity profiles also differs significantly. In the pressed TNT, a
fast increase of the conductivity up to a maximum value was observed with a following weak
decrease. For the cast TNT, the gradient of the increase is considerably smaller, and the value of
conductivity remains the same for about 1 $\mu$s. Using the notion of hot spots this can be
explained as following.

In pressed TNT, hot planes are produced at the contact of grains, and the reaction proceeds there.
In cast TNT at relatively low concentration of pores of the size near 0.1 mm or at the presence of
significantly larger pores, the time of burning until the overlap of reacting regions is
sufficiently higher. Not only the destruction of initial molecules but also the synthesis of main
reaction products have time to occur in the reacting regions. Thus, the matter reacted up to
different stages (from the initial one to fully reacted) is present in the cross-section where the
measurement is made. In such case, the magnitude of conductivity is limited by the amount of
condensed carbon which is observed. Hence, the course of the chemical reaction in cast and pressent
TNT is essentially different even at close detonation parameters. This is successfully detected by
the electric conductivity.

{\bf  Conclusions}

A new method was proposed for the diagnostics of the state of matter in the chemical peak region.
The method is based on the link between the magnitude of electric conductivity and the amount of
carbon condensed to nanostructures.

The spatial distribution of conductivity provides the real-time information by the tracing the
carbon transition from a non-conductive state to a conductive one.

The sensitivity of the diagnostics was demonstrated in such phenomena as the influence of the HE
grain size on the peak duration at powder density, the jet mechanism of the detonation propagation.
The role of hot spots in the kinetics of chemical reaction was demonstrated for TNT. For the first
time, it was shown directly that the course of chemical reactions in cast and pressed TNT differs
fundamentally.

The assumption on the absence of a direct relation between the chemical reaction zone and the
pressure was made based on the comparison of the data on the duration of pressure and conductivity
peaks obtained at the detonation of HMX, RDX and PETN. The pressure is not linked directly with the
reaction zone, it does not trace the state of matter in that zone, it is insensitive to the
chemical processes. Increase of the charge diameter and length results in the slowing down of the
rarefaction rather than in the increase of the duration of the reaction zone. The CJ point defined
as the point where the mass velocity is equal to the local speed of sound and the ending of the
chemical reaction zone can be separated in space. The detected weak influence of the density on the
chemical peak duration can be explained within the framework of the proposed hypothesis, and it
indicates that the cinetics is "detached"\ from mechanical parameters.

The diagnostics of the chemical peak by the electric conductivity opens wide possibilities for the
investigation of the reaction zone and for the development of a theory of hot spot ignition.
Presently, the conductivity method is a powerful tool for the investigation of the chemical
reaction zone at detonation. This method provides information on the course of chemical
transformations with high temporal resolution. It is suitable for the diagnostics of the matter
state in the chemical peak and for the investigation of non-classical and overdriven regimes.

{\bf Acknowledgements}

This work was supported by the Russian Foundation for Basic Research (project no. 15-03-01039a).

\vfill\eject

\end{document}